\documentclass[preprint,12pt,authoryear]{elsarticle}

\usepackage{amssymb}

\usepackage[colorlinks,allcolors=blue]{hyperref}

\usepackage{pdflscape}

\journal{Icarus}

\begin{document}

\begin{frontmatter}



\title{Oort cloud perturbations as a source of hyperbolic Earth impactors}


\author[inst1,inst2]{Eloy Peña-Asensio}
\author[inst3]{Jaakko Visuri}
\author[inst2,inst4]{Josep M. Trigo-Rodríguez}
\author[inst5,inst6]{Hector Socas-Navarro}
\author[inst7,inst8]{Maria Gritsevich}
\author[inst3]{Markku Siljama}
\author[inst1]{Albert Rimola}

\affiliation[inst1]{organization={Departament de Química, Universitat Autònoma de Barcelona},
            addressline={Carrer dels Til·lers}, 
            city={Bellaterra},
            postcode={08193}, 
            state={Catalonia},
            country={Spain}}

\affiliation[inst2]{organization={Institut de Ciències de l’Espai (ICE, CSIC)},
            addressline={Campus UAB, C/ de Can Magrans S/N}, 
            city={Cerdanyola del Vallès},
            postcode={08193}, 
            state={Catalonia},
            country={Spain}}

\affiliation[inst3]{organization={Finnish Fireball Network, Ursa Astronomical Association},
            addressline={Kopernikuksentie 1}, 
            city={Helsinki},
            postcode={00130}, 
            country={Finland}}
            
\affiliation[inst4]{organization={Institut  d’Estudis  Espacials  de  Catalunya  (IEEC)},
            addressline={Carrer del Gran Capità, 2}, 
            city={Barcelona},
            postcode={08034}, 
            state={Catalonia},
            country={Spain}}

\affiliation[inst5]{organization={Instituto de Astrofísica de Canarias},
            addressline={Avda Vía Láctea S/N}, 
            city={La Laguna},
            postcode={38205}, 
            state={Tenerife},
            country={Spain}}

\affiliation[inst6]{organization={Departamento de Astrofísica, Universidad de La Laguna},
            addressline={Avda Vía Láctea S/N}, 
            city={La Laguna},
            postcode={38205}, 
            state={Tenerife},
            country={Spain}}

\affiliation[inst7]{organization={Finnish Geospatial Research Institute},
            addressline={Vuorimiehentie 5}, 
            city={Espoo},
            postcode={02150}, 
            country={Finland}}

\affiliation[inst8]{organization={Faculty of Science, University of Helsinki},
            addressline={Gustaf Hällströmin katu 2a}, 
            city={Helsinki},
            postcode={00014}, 
            country={Finland}}

\begin{abstract}

The observation of interstellar objects 1I/'Oumuamua and 2I/Borisov suggests the existence of a larger population of smaller projectiles that impact our planet with unbound orbits. We analyze an asteroidal grazing meteor (FH1) recorded by the Finnish Fireball Network on October 23, 2022. FH1 displayed a likely hyperbolic orbit lying on the ecliptic plane with an estimated velocity excess of $\sim$0.7 km\,s$^{-1}$ at impact. FH1 may either be an interstellar object, indicating a high-strength bias in this population, or an Oort cloud object, which would reinforce migration-based solar system models. Furthermore, under the calculated uncertainties, FH1 could potentially be associated with the passage of Scholz's binary star system. Statistical evaluation of uncertainties in the CNEOS database and study of its hyperbolic fireballs reveals an anisotropic geocentric radiant distribution and low orbital inclinations, challenging the assumption of a randomly incoming interstellar population. Orbital integrations suggest that the event on March 9, 2017 (IM2) from CNEOS may have experienced gravitational perturbation during the Scholz fly-by, contingent upon velocity overestimation within the expected range. These findings suggest that apparent interstellar meteors may, in fact, be the result of accelerated meteoroid impacts caused by close encounters with massive objects within or passing through our solar system.

\end{abstract}




\end{frontmatter}


\section{Introduction}

In 2017, the Pan-STARRS1 telescope observed for the first time the reflected sunlight from a metric ($\sim$100 m) interstellar interloper, 1I/'Oumuamua \citep{Meech2017Natur}. Two years later, the second discovery of a large object (0.4--1 km) not gravitationally bound to the Sun, comet 2I/Borisov, was announced \citep{Guzik2020NatAs}. The discoverer of 2I/Borisov himself estimated that a spherical volume of 50 au radius may have 50 bodies of more than 50 meters in diameter \citep{BorisovShustov2021}. The Pan-STARRS survey's detection of 1I/'Oumuamua allows the calculation of a number density of 0.1 au$^{-3}$, corresponding to 10$^4$ similar objects within Neptune's orbit and an influx of 3 objects per day \citep{Jewitt2017ApJ}. By the expected power laws of object size distribution, a much more abundant population of smaller interstellar objects is expected to cross our solar system, which may eventually collide with the Earth. A review of interstellar objects and interlopers can be found in \citet{Jewitt20222023ARAA} and \citet{Seligman2023}.

When an object impacts the atmosphere at hypervelocity, friction with air particles progressively ablates the outer layers, radiating large amounts of energy \citep{Ceplecha1998SSRv, Silber2018AdSpR, Trigo2019hmep}. This luminous phase is known as a meteor or fireball, and its detection with ground-based or satellite instruments allows both the physicochemical analysis and the determination of the heliocentric orbit \citep{Jenniskens2009Natur, Trigo2006AG, Dmitriev2015PSS, Brown2016Icar, Devillepoix2019MNRAS, Borovi2020AA, Colas2020AA, Eloy2021MNRAS}. Recently, the first detections of interstellar meteors were claimed from the flashes spotted by the U.S. Department of Defense (DoD) satellite sensors and published on the NASA-JPL Center for NEOs Studies (CNEOS) website \citep{Tagliaferri1994hdtc}: the so-called IM1 occurred in 2014-01-08 \citep{Siraj2022ApJa} and IM2 in 2017-03-09 \citep{Siraj2022ApJb}, the latter being first identified as an interstellar candidate by \citet{Eloy2022AJ}.

In the early 20th century, the field of meteor science was predominantly focused on determining whether most meteors originated from interstellar or interplanetary sources \citep{Hughes1982VA}. However, it was not until the 1950s that the optical observations of fireballs generated by meteoroids exhibiting hyperbolic were reported \citep{Opik1950IrAJ, Almond1951MNRASI, Almond1952MNRASII}, in addition to subsequent meteor radar echoes detection of interstellar micrometeoroid impacts \citep{Weryk2004EMP, Froncisz2020PSS} and interstellar dust incoming flux measurements \citep{Meisel2002ApJa, Meisel2002ApJb}. Multiple automated meteor networks have detected numerous hyperbolic Earth impactors, most of which are pointed out as the result of the instrument and method limitations \citep{Stohl1970BAICz, Hajdukova2008EMP, Musci2012ApJ}. \citet{Hajdukova2020PSS} reported that, of the total number of recorded events, 12.5\% for CAMS, 11.9\% for SonotaCO, and 5.4\% for EDMOND were apparently hyperbolics. These events are clearly associated with low-quality detection and low angular elongation, so these large datasets cannot be used to discern hyperbolic impactors properly, and truly interstellar projectiles could remain hidden within the error bars. The identification of meteors with extra-solar provenance is a significant challenge and statements about the interstellar origin of IM1 and IM2 cannot be conclusive if the uncertainties of the data are not provided \citep{Vaubaillon2022JIMO}. Recent studies have even suggested that IM1 could be consistent with a common chondritic impactor assuming a lower atmospheric entry velocity \citep{Brown2023}.

\citet{Eloy2022AJ} and \citep{Brown2023} identified hyperbolic fireballs recorded by the United States Government (USG) satellite sensors, representing $\sim$1\% of total meter-sized impactors, events that are potentially meteorite-droppers. In contrast, there is no evidence of any recovered meteorite with a different composition from that of our solar nebula\footnote{On the other hand, whenever a meteorite is recovered it is attributed to our solar system by default.}. This fact opens several hypotheses: (1) CNEOS hyperbolic fireballs are spurious data; (2) There is a viable way for nearby stellar systems to be isotopically homogenous so extra-solar objects do not have distinctive non-chondritic elemental and isotopic compositions. The interstellar material exchange would be enough to smooth out any differences in the initial inventory of elements; (3) Incoming interstellar objects are biased towards low-strength properties and do not survive either the interstellar medium or the ablation process during the atmospheric entry; (4) There is an efficient mechanism by which objects that belong to our solar nebula acquire hyperbolic orbits. 

In this work, we present evidence supporting the latter hypothesis, assuming that the former remains unverified, a matter still awaiting clarification. We show that apparent interstellar meteors may actually be the result of accelerated projectile impacts due to gravitational perturbations induced by massive objects (stars, free-floating brown dwarfs, rogue planets, sub-stellar or sub-Jovian mass perturbers, primordial black holes...) shaping or visiting the outer part of our solar system. In particular, we analyze a likely hyperbolic asteroid-like grazing meteor recorded in Finland in 2022 exhibiting no deceleration, which could be associated with Scholz passage. Additionally, we discuss the IM2 hyperbolic fireballs of the CNEOS database, which may belong either to the Oort cloud or to a hypothetical Oort-like Scholz's cloud if its velocity is overestimated by 22\%. All events exhibit non-cometary compositions and probably are not of extra-solar provenance, which has profound implications for solar system formation models. In case they were truly interstellar in origin, the bias towards a high-strength composition of the incoming interstellar population would be reinforced.

\section{Methodology}\label{secMethods}

For the meteor science performed in this work, we use our verified Python pipeline \textit{3D-FireTOC} \citep{Eloy2021MNRAS, Eloy2021Astrodyn} which: performs the meteor positional reduction from the stellar astrometry accounting for asymmetric radial lens distortions \citep{Borovi1995} and atmospheric refraction by a revised Bennett's model \citep{Wilson2018PhDT}, employs the plane intersection method to reconstruct the atmospheric trajectory \citep{Ceplecha1987}, and computes the heliocentric orbit using the N-body orbital dynamics integrator \textit{REBOUND} and \textit{REBOUNDx} packages considering the gravitational harmonics (J2, J4) of the Earth and the Moon \citep{ReinSpiegel2015, Tamayo2020}. Uncertainties are calculated by generating 1,000 clones from the astrometry error fits assuming a normal distribution.

For mass estimation and event classification, it is necessary to calibrate the light curve. Using the visual magnitude of the same reference stars as in astrometry, we perform aperture photometry by subtracting the local background of each one. In this way, a logarithmic fit is conducted to relate pixel values with magnitudes. We correct the atmospheric extinction and calculate the absolute magnitude of the meteor (as observed at 100 km at the zenith).

The pre-atmospheric velocity is a critical quantity for orbit estimation and cannot be directly measured by optical devices. It is necessary to derive it from the distance traveled, for which a smooth function fit of the observed points is usually performed. For this purpose, we apply the function proposed by \citet{Whipple1957SCoA} that allows the velocity to be obtained straightforwardly from its derivative. However, for high-altitude grazing meteors, this model does not perform properly as it can not represent the velocity end. What is expected for the atmospheric entry of a meteoroid with these characteristics is a non-appreciable deceleration. For that reason, we assume that, within the error margins of the measurements, the pre-atmospheric and terminal velocities are virtually the same as a first approximation. Nevertheless, as we do not adjust the trajectory for the influence of gravity, we opt to analyze the initial third of the observed data points, conducting a linear fit with the mean value as the most likely velocity. The standard deviation of the fit serves as a measure of velocity uncertainty. As the entire trajectory can be used for velocity estimation without applying a deceleration model, this results in a smaller margin of error than expected for regular meteor velocity estimation from optical observations \citep{Egal2017Icar}.


Assuming the radiated energy by the meteor is proportional to the loss of kinetic energy in the form of mass loss \citep{Ceplecha1966BAICz}, which is only theoretically valid for atmospheric flight with no deceleration \citep{Gritsevich2011Icar}, the initial meteoroid mass can be computed from
\begin{equation}
m_0 = \int \frac{2}{\tau(v)v^2}I(t)dt,
\end{equation}

where $\tau$ is the luminous efficiency, $v$ is the velocity, $t$ is the time, $I = I_010^{-0.4M}$ is the radiated energy, $I_0 = 1,300\,\, W$ is the zero-magnitude radiant power for high-speed meteors \citep{Weryk2013PSS}, and $M$ the absolute magnitude. The luminous efficiency in percent is taken from \citet{Ceplecha1976JGR}: $\log\tau = -1.51 + \log v$ when $v$ $\geq$ 27 km\,s$^{-1}$. However, \citet{Borovicka2022AA} note that contemporary luminous efficiency models lead to $\sim$7 times less mass for velocities above 27 km\,s$^{-1}$, so our initial meteoroid mass may be overestimated by one order of magnitude. For example, using the \citet{Revelle2001ESASP} updated model where $\ln\tau = -1.53 + \ln v$ for $v$ $\geq$ 25.372 km\,s$^{-1}$, larger average luminous efficiency of are achieved.

Following \citet{Ceplecha1976JGR}, meteors can be classified according to the so-called $P_E$ criterion:
\begin{equation}
P_{E} = \log \rho_{e} - 0.42\log m_0 + 1.49\log v_{\infty} - 1.29\log \cos z_R,
\label{eq:ph_mass}
\end{equation}

where $\rho_{e}$ is the air density at terminal height, $v_{\infty}$ is the pre-atmospheric velocity, and $\cos z_R$ is the apparent radiant zenith distance. 

A more physical, dimensionless form of this criterion exists \citep{Moreno2020MNRAS}, however, as the analyzed event presents challenges in uniquely determining their atmospheric flight parameters, we turn to the original PE criterion form. From the classical third-order system describing the meteor body deceleration, numerous efforts have been made to define a height-velocity relation \citep{Kulakov1992AVest, Gritsevich2006SoSyR, Gritsevich2007SoSyR, Gritsevich2008SoSyR, Gritsevich2009AdSpR, Turchak2014JTAM, Lyytinen2016PSS, Sansom2019ApJ, Boaca2022ApJ, Eloy2023}. Following these works, the dynamics of a meteor can be characterized from the analytical solution using two dimensionless parameters, namely the ballistic coefficient $\alpha$ and the mass loss parameter $\beta$. It is possible to express $\alpha$ in terms of the meteoroid bulk density $\rho_m$, the pre-atmospheric shape factor $A_0$, the drag coefficient $c_d$, the atmospheric density at the sea level $\rho_0$, the height of the homogeneous atmosphere $h_0=7.16 \,\, km$, the meteoroid mass $m$, and the slope of the trajectory with the horizon $\gamma$:

 \begin{equation}\label{Eq:alpha}
\alpha = \frac{1}{2}\frac{c_d A_0 \rho_0 h_0}{m_0^{1/3} \rho_m^{2/3} \sin{\gamma}}. 
 \end{equation}

Assuming that the ablation of the body due to its rotation is uniform over the entire surface of the meteoroid \citep{Bouquet2014PSS}, the mass loss parameter can be calculated directly from the ablation coefficient $\sigma$ and the entry velocity:
 \begin{equation}
\beta = \frac{1}{6}\sigma v_{\infty}^2.
\end{equation}

We selected a uniformly distributed range of values for the ablation coefficient between 0.014 $s^2\,km^{-2}$ and 0.042 $s^2\,km^{-2}$ suitable for a rocky body based on both the classical single-body ablation and contemporary mass-loss models \citep{Ceplecha1998SSRv, Vida2018MNRAS}.

Due to the high altitudes and velocities of the atmospheric flight with an enhanced mass loss under the condition of minimal deceleration, standard dynamical fits, such as $\alpha$-$\beta$, may not perform properly as they are often organized as a function of velocity and are not primarily intended for high-height grazers, that is, for non-decelerating flights. Nevertheless, we can use the asymptotic form of the solution obtained to describe meteor trajectories at large values of the mass loss parameter given the formal fulfillment of the condition $\ln(2\alpha\beta)<h_e/h_0<\infty$, where $h_e$ is the end (terminal) height \citep{Stulov1997ApMRv, Stulov1998PSS, Gritsevich2008DokPh, Stulov2004PSS, Moreno2015Icar}.
To account for the possible change in velocity at the end of the luminous trajectory, we use the latest modification of this solution \citep{Moreno2015Icar, gritsevich2016approximating, Moreno2017ASSP}:
 \begin{equation}\label{Eq:v_e'}
v_{e'} = v_{\infty} \left( \frac{\ln(1-2\alpha\beta e^{-h_e/h_0})}{\beta} + 1 \right)^{1/2}.
\end{equation}

Using \textit{FireOwl} analysis software \citep{Visuri2020EPSC, Visuri2021LPICo2609V}, a Finnish Fireball Network tool that performs numerical integration of the meteoroid trajectory \citep{Moilanen2021MNRAS, Kyrylenko2023ApJ}, we recompute and contrast all the results. Finally, we check the dynamic association with some meteoroid stream or parent body by means of the well-known $D_{D}$ orbital dissimilarity criterion proposed by \citet{Drummond1981Icar}.

\section{Results}\label{secResults}

\begin{figure}[ht!]
\includegraphics[width=\linewidth]{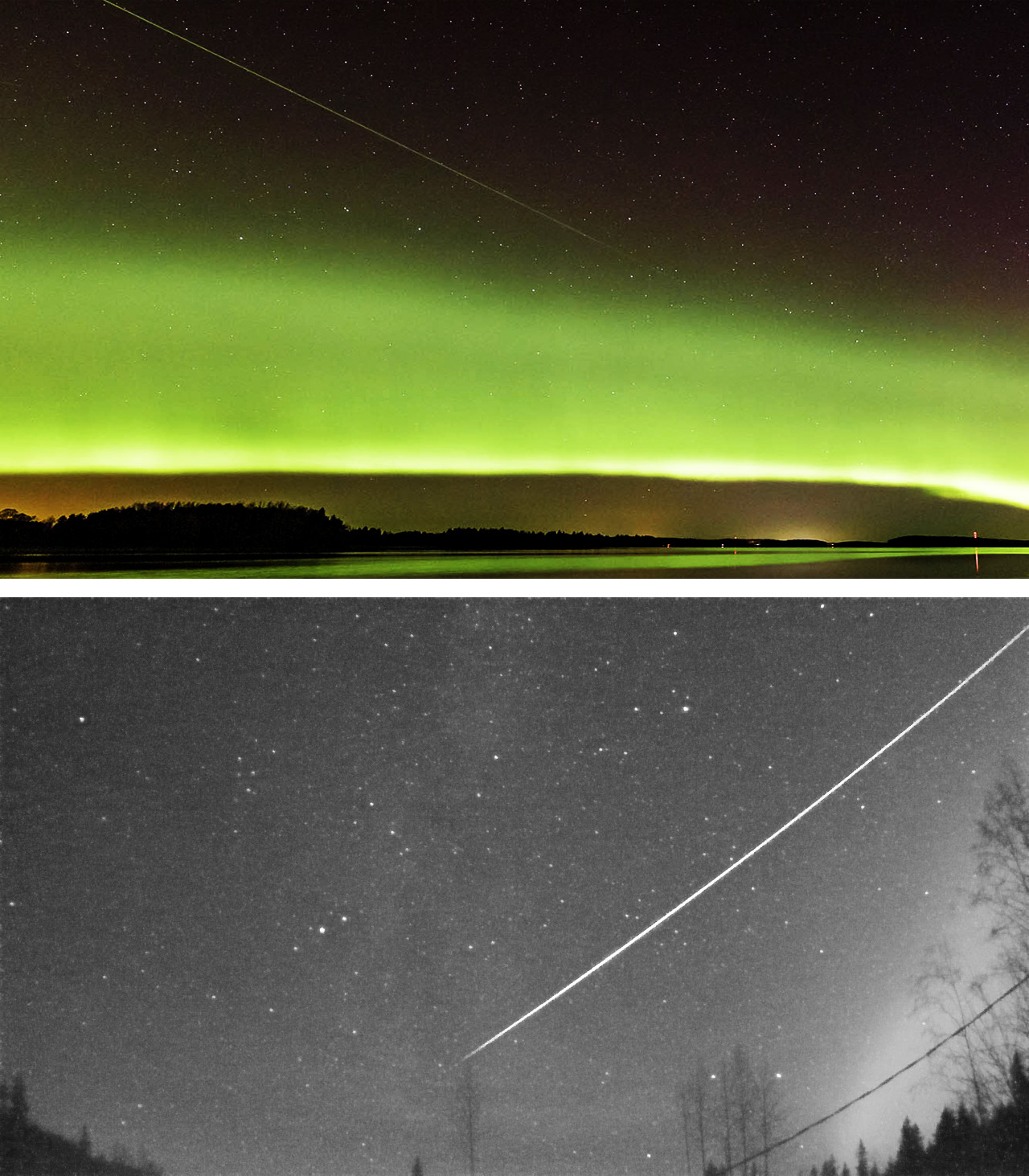}
\caption{Blended image of the FH1 videos recorded by Sastamala station with (top) and Nyrola station (bottom). The green illuminated area is an aurora borealis.
\label{fig:meteor_comp}}
\end{figure}

On October 23, 2022, at 19:38:34 (UTC), a very fast grazing meteor, hereafter FH1, flew through the sky of Finland and terminated over the Gulf of Bothnia. The event was observed by 3 stations of the Finnish Fireball Network (FFN) \citep{Gritsevich2014pim4, Trigo2015MNRAS, Lyytinen2016PSS, Visuri2021LPICo2609V, Moilanen2021MNRAS} and 1 image observation from the public: Nyrola (Sony IMX291; 1280x720 px; 4 mm f/0.95), Tampere (Hikvision DS-2CD2T87G2-L; 3840x2160 px; 2.8 mm f/1.0), Vaala (Watec 902H; 768x576 px; 3.8 mm f/0.8), and Sastamala (NIKON D750; 6016x4016 px; 14 mm f/4.0). These observations have been used in this study (Table \ref{tab:stations_det}). Figure \ref{fig:meteor_comp} shows two blended images of FH1 from the videos recorded by the Sastamala and Nyrola stations.

\begin{table}
\centering
\caption{Longitude, latitude, and altitude of the station recording the FH1 meteor.}
\label{tab:stations_det}
\begin{tabular}{lccc}
\hline
Station & Lon. ($^\circ$) & Lat. ($^\circ$) & Alt. (m.a.s.l.) \\
\hline
Nyrola & 25.5097 & 62.3425 & 203 \\
Vaala & 26.7317 & 64.3917 & 130 \\
Tampere & 23.7931 & 61.5119 & 171 \\
Sastamala & 23.0024 & 61.3826 & 63 \\
\hline
\end{tabular}
\end{table}

\begin{figure}[ht!]
\includegraphics[width=\linewidth]{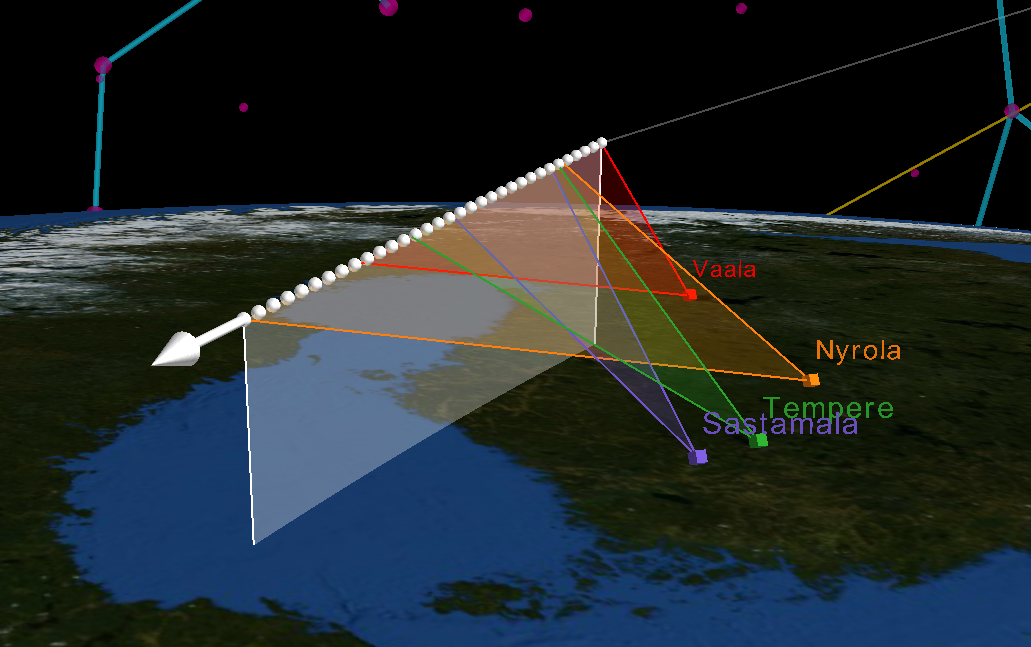}
\caption{3D scaled atmospheric flight reconstruction of the FH1 meteoroid by using the Python software \textit{3D-FireTOC}.
\label{fig:atm_traj}}
\end{figure}

The luminous phase of FH1 started at an altitude of 126.55$\pm$0.03 km (24.3104$\pm$0.0008$^\circ$ E, 63.6677$\pm$0.0002$^\circ$ N), traveling a distance of 409.47$\pm$0.09 km until ablation ended at 112.60$\pm$0.04 km altitude (18.558$\pm$0.002$^\circ$ E,\linebreak61.2358$\pm$0.0005$^\circ$ N). The flight angle with respect to the local horizon (i.e., the slope) was 3.588$\pm$0.013$^\circ$, with an azimuth of 229.915$\pm$0.007$^\circ$ (zero being north and positive in clockwise direction). Figure \ref{fig:atm_traj} shows the 3D scaled atmospheric flight reconstruction. The geocentric radiant, namely the corrected meteor anti-apex, is calculated outside of the gravitational influence field of the Earth and the Moon (at 10 times the Earth's Hill sphere), being the right ascension $\alpha_R$ = 117.160$\pm$0.009$^\circ$ and the declination $\delta_R$ = 19.444$\pm$0.020$^\circ$. The best convergence angle between the observations is $\sim$50$^\circ$ (for Sastamala and Vaala stations), where plane intersections with angles smaller than 5$^\circ$ are excluded (only for Sastamala and Tampere stations). Table \ref{tabA:ECEF_beg_end} in the Appendix shows the position vectors of the initial and final points of FH1's luminous path in the Earth-centered Earth-fixed coordinate system, as recorded by each of the four stations.

\begin{figure}[ht!]
\includegraphics[width=\linewidth]{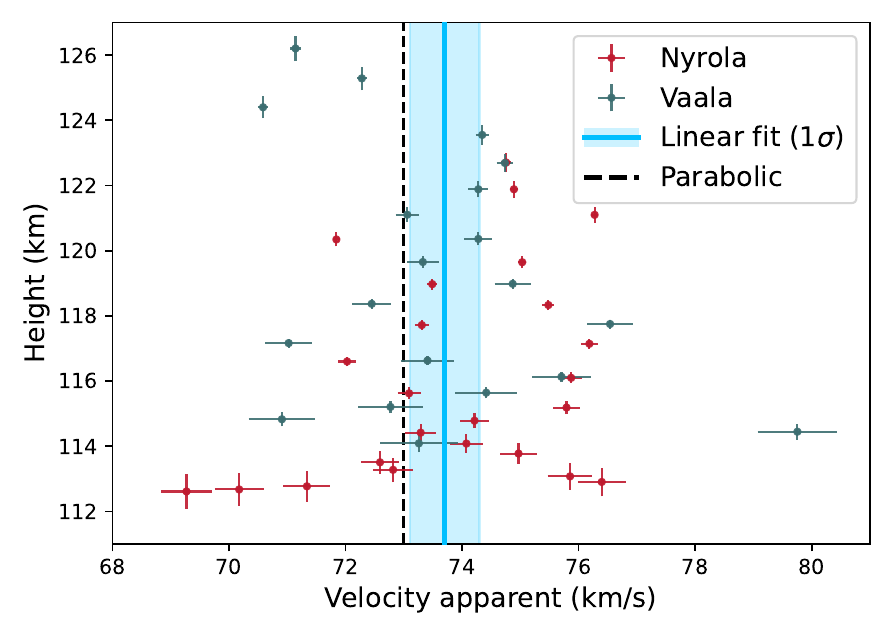}
\caption{Apparent velocity points of FH1 derived from Nyrola and Vaala observations computed at intervals of 0.2 seconds, fitted velocity, and the parabolic threshold. The error bars are multiplied by a factor of 10.
\label{fig:velocity}}
\end{figure}

Figure \ref{fig:velocity} shows the apparent point-to-point velocities, together with the fitted (73.7$\pm$0.6 km\,s$^{-1}$) and the parabolic velocity threshold for this specific atmospheric trajectory ($\sim$73 km\,s$^{-1}$). For 0.2 second intervals, Nyrola detection has 71\% of all instant velocity measurements above this threshold, while for Vaala it is 64\%. Nyrola and Vaala stations record at 25 fps, Tampere at 2.5, and Sastamala is an image of 5 seconds of exposure. Note that the apparent dispersion of the point-to-point velocities depends on the time interval selected and, paradoxically, the smaller the interval, the greater the dispersion, but the more accurate the final result. In the appendix, Tables \ref{tabA:NyrolaLocalObs} and \ref{tabA:VaalaLocalObs} offer the detected positions of FH1 for each frame, represented in a horizontal coordinate system comprising azimuth and elevation.


We are aware that the plane intersection method may not be appropriate for long-duration grazing events \citep{Borovi1992AA, Sansom2019Icar, Shober2020AJ}. Still, for this work, the approach described above is sufficient given the high velocity, brief duration (5.56 s), negligible atmospheric drag, and no dark flight calculations are required as no terminal mass is expected. Nonetheless, utilizing the fundamental equations of motion, we calculate the descent of an object due to gravitational acceleration under these conditions to be 147 m.

\begin{figure}[ht!]
\includegraphics[width=\linewidth]{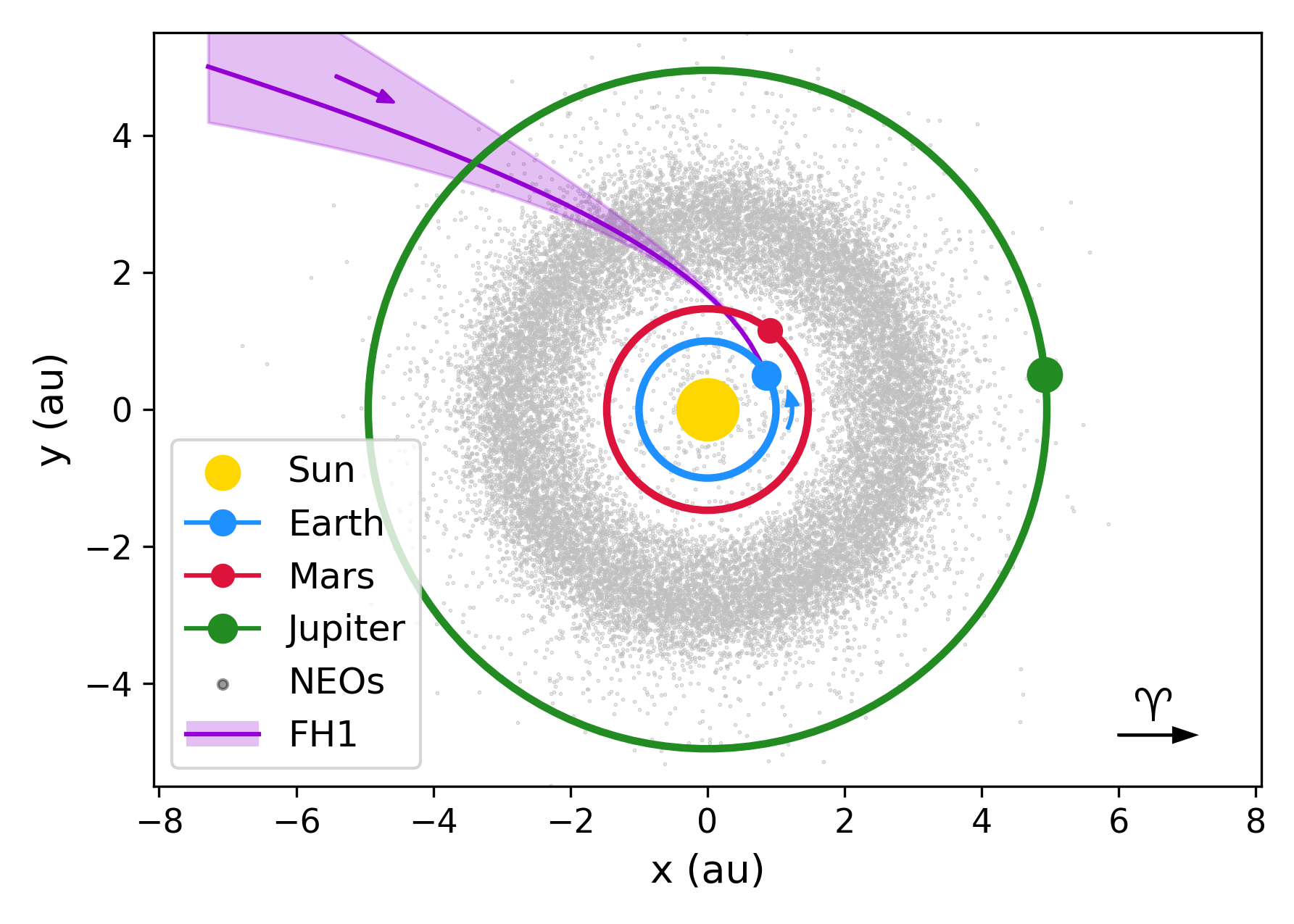}
\caption{Osculating heliocentric orbit of the FH1 meteoroid (J2000). The arrow at the bottom right shows the direction to the point of the vernal equinox.
\label{fig:orbit}}
\end{figure}

From the inbound velocity, we estimate the heliocentric osculating orbital elements at impact to be the following: semi-major axis $a$ = $-$8$\pm$5 au, eccentricity $e$ = 1.07$\pm$0.06, inclination $i$ = 177.18$\pm$0.04$^\circ$, the longitude of the ascending node $\Omega$ = 30.10390$\pm$0.00010$^\circ$, argument of periapsis $\omega$ = 16.1$\pm$0.8$^\circ$, and true anomaly $f$ = 343.9$\pm$0.8$^\circ$. The orbit shows no close encounters with any planets. Figure \ref{fig:orbit} illustrates the obtained heliocentric hyperbolic orbit.

\begin{figure}[ht!]
\includegraphics[width=\linewidth]{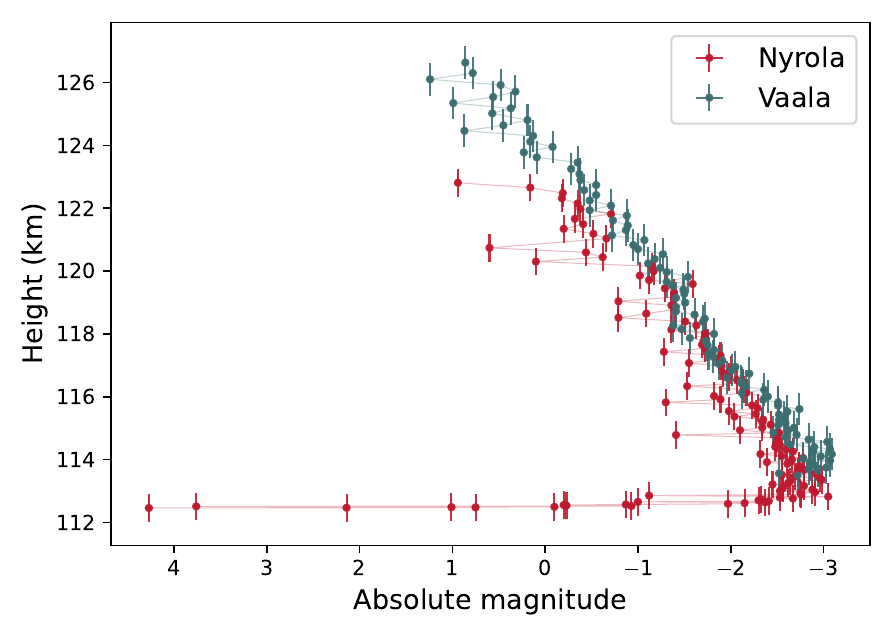}
\caption{Photometry of FH1 from Nyrola and Vaala stations. The mean uncertainty of the magnitude is 0.44 and 0.55 respectively.
\label{fig:photometry}}
\end{figure}

Figure \ref{fig:photometry} shows the meteoroid absolute magnitude for every frame from Nyrola and Vaala stations, which are in good agreement with each other. FH1 curve light has a mean luminous efficiency of $\tau = 2.278\pm0.018\,\, \%$, a peak brightness of $M = -3.0\pm0.5$, and yields a photometric initial mass of $m = 1,312\pm54 \,\,g$. Using the \citet{Revelle2001ESASP} updated model yields an average luminous efficiency of 15.96$\pm$0.13 \%. Note that the meteoroid underwent a smooth and gradual ablation without any flares or catastrophic disruption, resulting in the absence of saturated pixels in all recordings.

For FH1 meteor we obtain $P_E = -4.173\pm0.009$. \citet{Ceplecha1976JGR} classified meteors as type I when $P_E$ $>$ -4.6, which \citet{Ceplecha1998SSRv} assigned to ordinary chondrites. Note that the photometric mass used for $P_E$ classification must be computed using luminous efficiency from \citet{Ceplecha1976JGR} in Eq. \ref{eq:ph_mass}. The obtained value is in good agreement with our estimated luminous efficiency as \citet{Revelle2001ESASP} found that asteroidal fireballs should be around 5.57\% and 1.35\% for carbonaceous chondrite-like fireballs. This is consistent also with similar works \citet{Subasinghe2017PSS, Drolshagen2021AAa, Drolshagen2021AAb}. In any case, the FH1 meteoroid has a consistency equal to or greater than ordinary chondrites, tending towards high-strength materials. 

Conservatively, taking an asteroid-like bulk density of 3,700 $kg\,m^{-3}$, we compute an initial meteoroid diameter of 8.75$\pm$0.12 cm, which in contrast may be 4.59$\pm$0.06 cm in diameter based on modern luminous efficiency models. Given the inferred meteoroid size and bulk density, an asteroidal origin seems likely \citep{Blum2006ApJ, Trigo2006MNRAS}, although it could also be compatible with rocky pebbles ejected by cometary disintegration during inner solar system trips \citep{Trigo2022MNRAS}. From Eq. \ref{Eq:alpha}, we obtain a value of $\alpha$ of 444$\pm$4 or 849$\pm$8 depending on the initial photometric mass estimates previously calculated, and $\beta=25\pm7 $, far away from being a meteorite-dropper event \citep{Gritsevich2012CosRe, Sansom2019ApJ, Boaca2022ApJ}. Eq. \ref{Eq:v_e'} yields a velocity decrease over 5 m\,s$^{-1}$ and 10 m\,s$^{-1}$, which is below the resolution of the measurements and within the uncertainty margin estimated for the velocity along the flight.

We corroborate with the \textit{FireOwl} pipeline that an asteroid-like meteoroid with no catastrophic disruption and the estimated characteristics would behave in agreement with the observations. However, differences between compositions are almost marginal as the projectile experiences a low air drag during its $\sim$5.56 seconds of flight. Therefore, on this occasion, the dynamic models cannot provide conclusive results concerning the meteoroid density. 

The two candidates are the Taurids swarm and the comet P/2015 A3, with a $D_{D}$=0.160 and $D_{D}$=0.177, respectively. These values are well above the typically accepted threshold \citep{Galligan2001MNRAS}, so this event is definitely not associated with any known parent body or meteoroid stream.

In summary, FH1 was a non-cometary centimeter-sized meteoroid in an inbound retrograde likely hyperbolic orbit lying almost on the ecliptic plane. It exhibits no close encounter with any known planet and a velocity excess at impact of $\sim$0.7 km\,s$^{-1}$ with respect to the barycentre of the solar system. Photometry of the meteor phase yields an asteroid-like (or higher) bulk density. FH1 is the first likely hyperbolic event detected by the FFN since the beginning of the year 2004, with over 2,000 manually analyzed meteors. All computed parameters can be found in Table \ref{tab:GrazerParameters}.

\begin{table}
\centering
\footnotesize
\caption{Atmospheric flight, photometric, physical, and heliocentric orbital (J2000) computed parameters of FH1 grazing meteor. The values with two results correspond to the luminous efficiency models considered: \citet{Ceplecha1976JGR} on the left and \citet{Revelle2001ESASP} on the right.  \label{tab:GrazerParameters}}
\begin{tabular}{lcc}
\hline
\textbf{Parameter} &  & \textbf{Value} \\
\hline
Reference time (UTC)                & $t_0$                         & 2022-10-23 19:38:34 \\
Velocity (km\,s$^{-1}$)             & $v_{\infty}$, $v_0$, $v_e$    & 73.7$\pm$0.6 \\
Initial latitude ($^\circ$)         & $\varphi_0$                   & 63.6677$\pm$0.0002 N \\
Initial longitude ($^\circ$)        & $\lambda_0$                   & 24.3104$\pm$0.0008 E  \\
Initial height (km)                 & $h_0$                         & 126.55$\pm$0.03 \\
Final latitude ($^\circ$)           & $\varphi_e$                   & 61.2358$\pm$0.0005 N \\
Final longitude ($^\circ$)          & $\lambda_e$                   & 18.558$\pm$0.002 E \\
Final height (km)                   & $h_e$                         & 112.60$\pm$0.04 \\
Duration (s)                        & $\Delta t$                    & 5.56 \\
Length (km)                         & $\Delta l$                    & 409.47$\pm$0.09 \\
Slope ($^\circ$)                    & $\gamma$                      & 3.588$\pm$0.013 \\
Azimuth ($^\circ$)                  & $A$                           & 229.915$\pm$0.007 \\
Peak brightness                     & $M$                           & -3.0$\pm$0.5 \\
Luminous efficiency (\%)            & $\tau$                        & 2.278$\pm$0.018 $\mid$ 15.96$\pm$0.13 \\
Ablation coef. ($s^2\,km^{-2}$)     & $\sigma$                      & 0.014 $\mid$ 0.042 \\
Initial phot. mass (g)              & $m_0$                         & 1,312$\pm$54 $\mid$ 187$\pm$7 \\
P$_E$ criterion                     & $P_E$                         & -4.173$\pm$0.009 \\
Meteoroid density (kg\,m$^{-3}$)    & $\rho_m$                      & 3,700 \\
Initial diameter (cm)               & $D$                           & 8.75$\pm$0.12 $\mid$ 4.59$\pm$0.06 \\ 
Geo. velocity (km\,s$^{-1}$)        & $v_R$                         & 72.7$\pm$0.6 \\
Geo. radiant (RA) ($^\circ$)        & $\alpha_R$                    & 117.160$\pm$0.009 \\
Geo. radiant (Dec) ($^\circ$)       & $\delta_R$                    & 19.444$\pm$0.020 \\
Hyp. excess (km\,s$^{-1}$)          & $\Delta_v$                    & $\sim$0.7 \\
Hel. velocity (km\,s$^{-1}$)        & $v_H$                         & 43.0$\pm$0.6 \\
Semi-major axis (au)                & $a$                           & -8$\pm$5\\
Eccentricity                        & $e$                           & 1.07$\pm$0.06 \\
Inclination ($^\circ$)              & $i$                           & 177.18$\pm$0.04 \\
Long. of the asc. node ($^\circ$)   & $\Omega$                      & 30.10390$\pm$0.00010 \\
Argument of periapsis ($^\circ$)    & $\omega$                      & 16.1$\pm$0.8 \\
Periapsis distance (au)             & $q$                           & 0.9748 $\pm$0.0008 \\
True anomaly ($^\circ$)             & $f$                           & 343.9$\pm$0.8 \\
Ballistic coef.                     & $\alpha$                      & 444$\pm$4 $\mid$ 849$\pm$8 \\
Mass-loss parameter                 & $\beta$                       & 25$\pm$7 \\
Deceleration (m\,s$^{-2}$)          & a                             & 0.87$\pm$0.01 $\mid$ 1.67$\pm$0.02 \\
\hline
\end{tabular}
\end{table}

\section{Discussion}

The distinctive attributes of FH1, primarily its remarkably high eccentricity, could conceivably prompt conjectures regarding its interstellar origin. Notably, such speculations have been posited recently in the context of certain hyperbolic fireball events cataloged in the CNEOS database. However, a critical observation emerges when analyzing orbital inclination, which appears as a key indicator that urges to exercise caution before leaping to interstellar suppositions. We need first to investigate more plausible scenarios, including the possibility that these intriguing projectiles are either indigenous to our solar system, subject to measurement inaccuracies, or potentially subjected to gravitational accelerations.

In this section, due to the similarity with FH1, we discuss the possible interstellar origin of some CNEOS fireballs considering their uncertainties from events detected independently by ground-based stations. Additionally, we put forth the hypothesis that hyperbolic Earth impactors may be celestial bodies native to our solar nebula, which have been perturbed by close encounters with massive objects. More precisely, we propose that IM2's trajectory aligns exceptionally well in time and direction with the Scholz system fly-by when considering an overestimated velocity.

\subsection{CNEOS 'interstellar' fireballs} \label{sec:CNEOS}

As of October 2023, the CNEOS public database includes $\sim$956 fireballs starting from 1988. Among them, 6 events have hyperbolic orbits (see Table \ref{tab:cneos_hyp}). These interstellar candidates have orbital inclinations lower than 25$^\circ$ (with an average of 12$\pm$9$^\circ$). As interstellar interlopers may originate from any part of the sky, the expected inclination should be an isotropic probability density function, which follows a sinusoidal distribution \citep{Engelhardt2017AJ} and, therefore, is uniform in $\cos i$. This implies that the random likelihood that $n$ orbital inclinations fulfill $\mid i \mid \leq \theta ^\circ$ where $i \in [-\pi/2,\pi/2]$ is $(1-\cos\theta)^{n}$. Consequently, the orbits of 1I/'Oumuamua and 2I/Borisov had a likelihood of being lower than $\mid-58^\circ\mid$ (the largest inclination which is 1I/Omumuamu's) of $\sim$22\%. By comparison, the likelihood of detecting six interstellar objects with inclination orbits smaller than 25$^\circ$ is $\sim$0.00007\%. This is without considering that all 6 events are in prograde orbits, which should be expected in the 50\% of extra-solar visitors and would further reduce the likelihood. Therefore, multiple options can be inferred: there are shortcoming data in the CNEOS database, these hyperbolic fireballs belonged to our solar system, or they came from sources with a directional bias. 

\begin{table}
\centering
\footnotesize
\caption{CNEOS hyperbolic fireballs with geocentric radiant, heliocentric velocity, semi-major axis, eccentricity, and orbital inclination.}
\label{tab:cneos_hyp}
\begin{tabular}{lcccccc}
\hline
Date (UTC) & $\alpha_R$ ($^\circ$) & $\delta_R$ ($^\circ$) & V$_h$ (km\,s$^{-1}$) & a (au) & e & i ($^\circ$) \\
\hline
2022-07-28 01:36:08 & 276.5 & 14.8 & 46.9 & -1.98 & 1.44 & 23.47 \\
2021-05-06 05:54:27 & 62.5 & 12.2 & 44.1 & -2.64 & 1.15 & 6.05 \\
2017-03-09 04:16:37 (IM2) & 170.6 & 34.1 & 50.1 & -1.22 & 1.57 & 24.03 \\
2015-02-17 13:19:50 & 339.3 & -9.6 & 44.0 & -1.45 & 1.10 & 1.12 \\
2014-01-08 17:05:34 (IM1) & 88.9 & 13.3 & 61.1 & -0.46 & 2.42 & 10.05 \\
2009-04-10 18:42:45 & 107.8 & 4.5 & 45.5 & -1.91 & 1.33 & 6.52 \\
\hline
\end{tabular}
\end{table}

As the error bars are not provided by USG sensors, it is necessary to narrow down the uncertainties and determine the frequency of spurious data in the database. \citet{Devillepoix2019MNRAS} reported that CNEOS fireball radiants are off for most events, sometimes by only a couple of degrees but other times as much as 90$^\circ$. They compared the radiants of 9 events recorded simultaneously by ground-based stations and found that the velocity vector of 4 of them was incorrectly measured by the USG sensors: Buzzard Coulee (2008-11-21 00:26:44), 2008 TC3 (2008-10-07 02:45:45), DN150102 - Kalabity (2015-01-02 13:39:11), and Crawford Bay (2017-09-05 05:11:27). As we calculate a different radiant for Crawford Bay event based on CNEOS data and, as pointed in \citet{Eloy2022AJ}, 2008 TC3 was missing a minus sign in the z velocity component, we decide to recompute the mean radiant position and velocity deviations of CNEOS fireballs including also the recent independently analyzed events Sariçiçek (2015-09-02 20:10:30) \citep{Unsalan2019MPS}, Ozerki (2018-06-21 01:16:20) \citep{Maksimova2020MPS, Kartashova2020PSS}, Viñales (2019-02-01 18:17:10) \citep{Zuluaga2019MNRAS}, 2019 MO (2019-06-22 21:25:48) \citep{NASAJPLHorizon}, Flensburg (2019-09-12 12:49:48) \citep{Borovicka2021MPS}, Novo Mesto (2020-02-28 09:30:34) \citep{Vida2021EPSC}, Ådalen (2020-11-07 21:27:04) \citep{Kyrylenko2023ApJ}, and 2022 EB5 (2022-03-11 21:22:46) \citep{NASAJPLHorizon}.

\begin{landscape}
\begin{table}
\centering
\scriptsize
\caption{Comparison of 17 fireballs detected by USG sensors and published on CNEOS website with independent ground-based analysis. The geocentric radiant in right ascension and declination, the entry velocity, the radiant position angle deviation, and the velocity deviation are shown. The papers used as a reference are listed in the last column. Some of the geocentric parameters have been calculated from the apparent atmospheric data.}
\label{tab:CNEOSdeviations}
\begin{tabular}{lcccccccccc}
\hline
Event & Date & $\alpha_R^{USG}$ & $\delta_R^{USG}$ & $v_e^{USG}$ & $\alpha_R^{REF}$ & $\delta_R^{REF}$ & $v_e^{REF}$ & $\Delta_R$ & $\Delta v_e$ & Reference \\
 & (UTC) & ($^\circ$) & ($^\circ$) & (km\,s$^{-1}$) & ($^\circ$) & ($^\circ$) & (km\,s$^{-1}$) & ($^\circ$) & (\%) & 
\\ \hline
2008 TC3 & 2008-10-07 02:45:45 & 351.6 & 9.06 & 13.3 & 316.8 & 7.2 & 12.42 & 34.49 & 7.09 & \citet{Scheirich2010MPS} \\
Buzzard Coulee & 2008-11-21 00:26:40 & 184.8 & 50.8 & 12.9 & 300 & 75 & 18.6 & 47.24 & 30.65 & \citet{Milley2010MsT} \\
Kosice & 2010-02-28 22:24:47 & 114.5 & 33.5 & 15.1 & 114.3 & 29 & 15 & 4.5 & 0.67 & \citet{Borovi2013MPS} \\
Chelyabinsk & 2013-02-15 03:20:21 & 334.5 & -1.5 & 18.6 & 328.28 & 0.28 & 19.03 & 6.47 & 2.26 & \citet{Borovi2013Nature} \\
Kalabity & 2015-01-02 13:39:11 & 53.8 & 33.5 & 18.1 & 64.3 & 51.7 & 15.4 & 19.73 & 17.53 & \citet{Devillepoix2019MNRAS} \\
Romania & 2015-01-07 01:05:59 & 118.7 & 6.1 & 35.7 & 113.8 & 10.13 & 27.76 & 6.31 & 28.6 & \citet{Borovi2017PSS} \\
Sariçiçek & 2015-09-02 20:10:30 & 61.1 & 45.2 & 21.1 & 264.8 & 59.4 & 13.0 & 73.6 & 62.31 & \citet{Unsalan2019MPS} \\
Baird Bay & 2017-06-30 14:26:45 & 273.6 & -16.1 & 15.2 & 272.14 & -12.5 & 15.1 & 3.87 & 0.66 & \citet{Devillepoix2019MNRAS} \\
Crawford Bay & 2017-09-05 05:11:27 & 203.7 & 1.8 & 14.7 & 205.12 & 3.13 & 16.5 & 1.94 & 10.91 & \citet{Hildebrand2018LPI} \\
2018 LA & 2018-06-02 16:44:12 & 248.7 & -9.7 & 11.8 & 244.19 & -10.32 & 12.375 & 4.48 & 4.65 & \citet{Jenniskens2021MAPS} \\
Ozerki & 2018-06-21 01:16:20 & 310.6 & 44.5 & 14.4 & 307.51 & 43.11 & 14.9 & 2.63 & 3.36 & \citet{Kartashova2020PSS} \\
Viñales & 2019-02-01 18:17:10 & 325.4 & -42.8 & 16.3 & 324.72 & -41.43 & 16.9 & 1.46 & 3.55 & \citet{Zuluaga2019MNRAS} \\
2019 MO & 2019-06-22 21:25:48 & 217.2 & -16.1 & 9.6 & 237.3 & -15.6 & 9.3 & 19.33 & 3.23 & \citet{NASAJPLHorizon} \\
Flensburg & 2019-09-12 12:49:48 & 183.1 & -21.3 & 14.6 & 183.5 & -18.55 & 14.77 & 2.78 & 1.15 & \citet{Borovicka2021MPS} \\
Novo Mesto & 2020-02-28 09:30:34 & 332 & 2.0 & 21.5 & 330.92 & 2.32 & 22.098 & 1.13 & 2.71 & \citet{Vida2021EPSC} \\
Ådalen & 2020-11-07 21:27:04 & 359.2 & 47.3 & 12.4 & 358.1 & 47.6 & 13.5 & 0.8 & 8.15 & \citet{Kyrylenko2023ApJ} \\
2022 EB5 & 2022-03-11 21:22:46 & 157.5 & 38.1 & 13.2 & 157.0 & 37.6 & 13.0 & 0.64 & 1.54 & \citet{NASAJPLHorizon} \\
\hline
\end{tabular}
\end{table}
\end{landscape}

From the comparison of independently analyzed events presented in Table \ref{tab:CNEOSdeviations}, it is found that the CNEOS radiants have a mean deviation of 21.3$\pm$43.2$^\circ$ in right ascension and 4.8$\pm$6.8$^\circ$ in declination, and a mean entry velocity deviation of 11.1$\pm$15.6\%. It can be deduced from the standard deviations that the errors do not follow a normal distribution, as these distributions strongly depend on the relative geometry of the sensor and the fireball. 

Eliminating the fireballs that appear as outliers based on the radiant position and velocity errors (2008 TC3, Buzzard Coulee, Kalabity, Romania, Sariçiçek, 2019 MO), deviations become Gaussian where the radiant would be reduced to 1.9$^\circ$ in right ascension and 1.7$^\circ$ in declination, and a 3.6\% deviation in velocity, which is the scenario we assume for the study of the CNEOS fireballs. Therefore, this assumption remains valid solely under the condition that the hyperbolic CNEOS events are part of the same distribution as the events characterized by elliptical orbits in Table \ref{tab:CNEOSdeviations}. Hence, if these events are indeed outliers, the presented results here should be ignored. Consequently, 65\% of CNEOS events provide measurements accurate enough for a rough estimation of heliocentric orbits. However, this also implies that the results for two out of the six hyperbolic fireballs will be inaccurate, rendering these estimated errors inapplicable. Note that there is a lack of a significant correlation between the velocity error and the actual velocity value. When a linear fit is performed, it yields a coefficient of determination as low as 0.027. In principle, we should not assume that the faster hyperbolic ones will necessarily exhibit larger errors

\begin{figure}
\includegraphics[width=\linewidth]{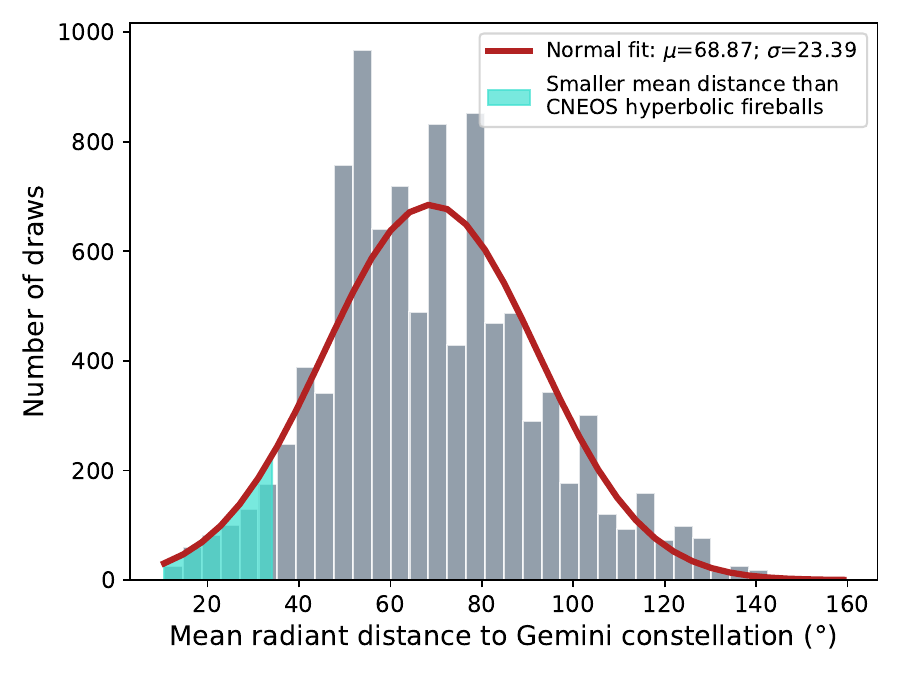}
\caption{Result of randomly selecting 6 CNEOS events and computing the mean minimum distance to the Gemini  constellation center of 4 of them. The calculation is repeated 10,000 times. In cyan color are shown the draws with equal or smaller distances than the 6 CNEOS hyperbolic events. The fit to a Gaussian distribution is shown in red.
\label{fig:draws}}
\end{figure}

These hyperbolic events, 2009-04-10 18:42:45, 2014-01-08 17:05:34 (IM1), 2017-03-09 04:16:37 (IM2), and 2021-05-06 05:54:27, appear to be somehow clusterized. The geocentric radiants of these 4 fireballs are suspiciously distributed around the Gemini constellation with an average radiant distance to the constellation center point of 34.2$^\circ$. We check the likelihood that 4 out of 6 randomly selected events from the CNEOS database (the 255 fireballs analyzed in \citet{Eloy2022AJ}) have a lower mean distance value to Gemini (10,000 draws). We find that these 4 hyperbolic events represent 1.7$\sigma$ with respect to the mean, which denotes a probability of 6.9\% having occurred by chance (see Figure \ref{fig:draws}). From a completely isotropic radiant distribution, the probability of having obtained a smaller distance for 4/6 events is 4.3\%. However, this does not strictly imply they are associated with each other as the anisotropic radiant distribution could be explained as well by both observational bias and solar system induced dynamics. 

\subsection{The massive objects fly-by hypothesis} \label{sec:Scholz}

Not all gravitationally unbound objects from our solar system are necessarily interstellar interlopers. There are different mechanisms capable of accelerating objects native to our solar nebula into hyperbolic orbits. Some of them are the secular perturbations induced by the Galactic disk or fly-by impulsive interaction of massive extra-solar bodies \citep{Fouchard2011AA, Krolikowska2017MNRAS}. Furthermore, close encounters with the Sun or giant planets may result in an inbound excess velocity, although these are not frequent enough to explain the observed hyperbolic meteor orbits \citep{Hajdukova2014MPS, Hajdukova2019msme}. Mercury has also been identified as a possible efficient producer of hyperbolic projectiles to Earth \citep{Wiegert2014Icar}. Other exotic hypotheses suggest unseen stellar companions to the Sun \citep{Davis1984Natur} or unknown planets \citep{HectorSocas2022arXiv} as a source of hyperbolic Earth impactors.

When the idea of the Oort cloud was introduced, i.e. a very distant region with long-period comets, it was also pointed out the existence of a mechanism to shorten their perihelia, for example, inbound hyperbolic injection produced by passing stars \citep{Oort1950BAN}. Indeed, recent studies suggest that stellar close encounters send accelerated bodies into the planetary zone \citep{Dybczy2022AA}. \citet{Higuchi2020MNRAS} showed that celestial bodies of sub-stellar mass (down to approximately 0.2 Jupiter masses) possess the ability to divert Oort cloud comets into hyperbolic trajectories characterized by small eccentricity but large perihelion distance. Other stellar systems may have also their own Oort-like clouds which could induce an influx of extra-solar objects through the planetary region when approaching the Sun \citep{Stern1987Icar}. The most recent stellar fly-by to our solar system was the low-mass binary star WISE J072003.20-084651.2, also known as Scholz's star (hereafter Scholz), which crossed the outer layers of the Oort cloud at 52$_{-14}^{+23}$ kau about 70$_{-10}^{+15}$ kya ago \citep{Scholz2014AA, Burgasser2015AJ, Mamajek2015ApJ}.

\citet{delaFuente2018MNRAS} analyzed hyperbolic small bodies of the data provided by JPL’s solar system Dynamics Group Small-Body and the Minor Planets Center (MPC) databases. They found strong anisotropies on the geocentric radiant distribution with a statistically significant overdensity of high-speed radiants towards the constellation of Gemini, which appears to be consistent in terms of time and location with the Scholz fly-by. Precisely the geocentric radiant of FH1 falls in this constellation, as well as 4 of the 6 hyperbolic fireballs from the CNEOS database that are close to Gemini or the recent Scholz motion direction.

\begin{figure*}[ht!]
    \includegraphics[width=\linewidth]{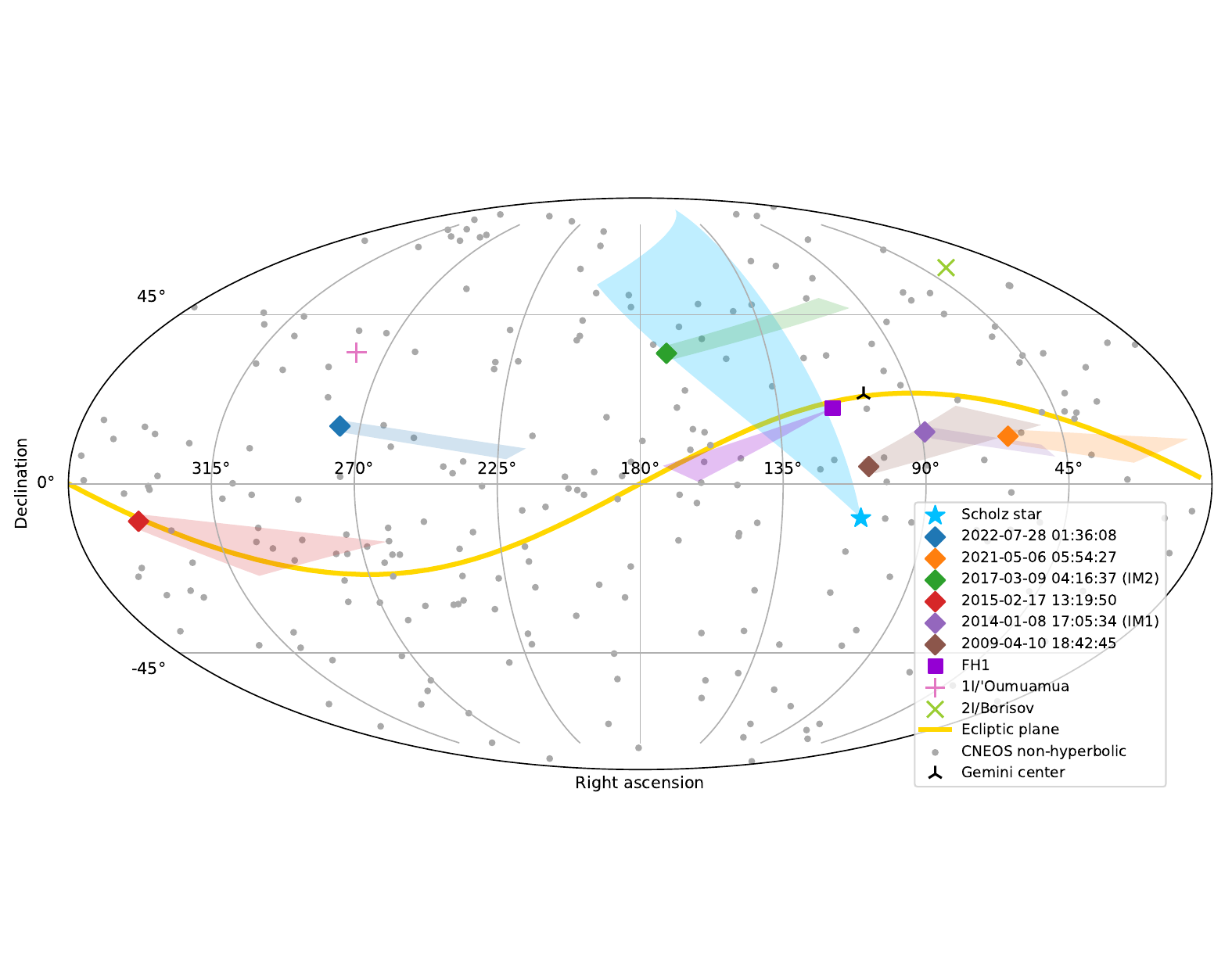}
    \caption{Apparent motion of the objects starting from their geocentric radiants during the backward orbital integration. Scholz, FH1 grazer event, 1I/'Oumuamua \citep{Meech2017Natur}, and 2I/Borisov \citep{JuliaDeLeon2020MNRAS} are shown. The 6 dated events correspond to the hyperbolic fireballs in the CNEOS database and include the new mean deviation found for the 17 fireballs compared. All CNEOS non-hyperbolic events are also depicted, together with a center point of the Gemini constellation. Markers represent the radiant position at impact or at the current time in the case of Scholz. The ecliptic plane is plotted in yellow.
    \label{fig:radiants}}
    \end{figure*}

We test the compatibility of this hypothesis by integrating backward in time the FH1 grazer and the 6 CNEOS interstellar candidates for 109,000 years to account for the estimated upper time limit of the Scholz close encounter (85,000 years). To this end, we use an orbital integrator based on a leapfrog scheme with different time steps to properly resolve the Earth-induced zenith attraction prior to the impact \citep{HectorSocas2019AA}. We account for the gravitational influence of the Sun, Earth, Moon, Mars, Jupiter, Saturn, Uranus, and Neptune by querying ephemerids to JPL HORIZONS system. Figure \ref{fig:radiants} shows the apparent motion of the objects starting from their geocentric radiants for the considered time. None of the events experienced a close encounter with planets, only 2021-05-06 05:54:27 fireball passed $\sim$5 years ago at $\sim$3 times Hill radius from Uranus.

New results definitively dismiss the possibility that Scholz may have penetrated the dynamically active inner Oort cloud region ($<$20 kAU), but support the notion that it would have passed through the outer Oort cloud where objects can have stable orbits \citep{Dupuy2019AJ, delaFuente2022RNAAS}. Given the apparently better constrained time ($\sim$80 kya) and distance ($\sim$68 kau) for the Scholz close encounter, and its current separation from the Sun (6.80 pc), it can be computed a linear velocity with respect to the solar system barycenter of $v^*=82.4\pm0.3\, km\,s^{-1}$, which is a valid approximation for 100 kya within 2.5\% accuracy \citep{Mamajek2015ApJ}. Considering the fly-by occurring at high velocity and the low mass of the Scholz system ($M_*=165\pm7\, M_{Jup}$), it appears plausible that small bodies may have been injected towards the Earth.

\begin{figure}
\centering
\includegraphics[width=1\linewidth]{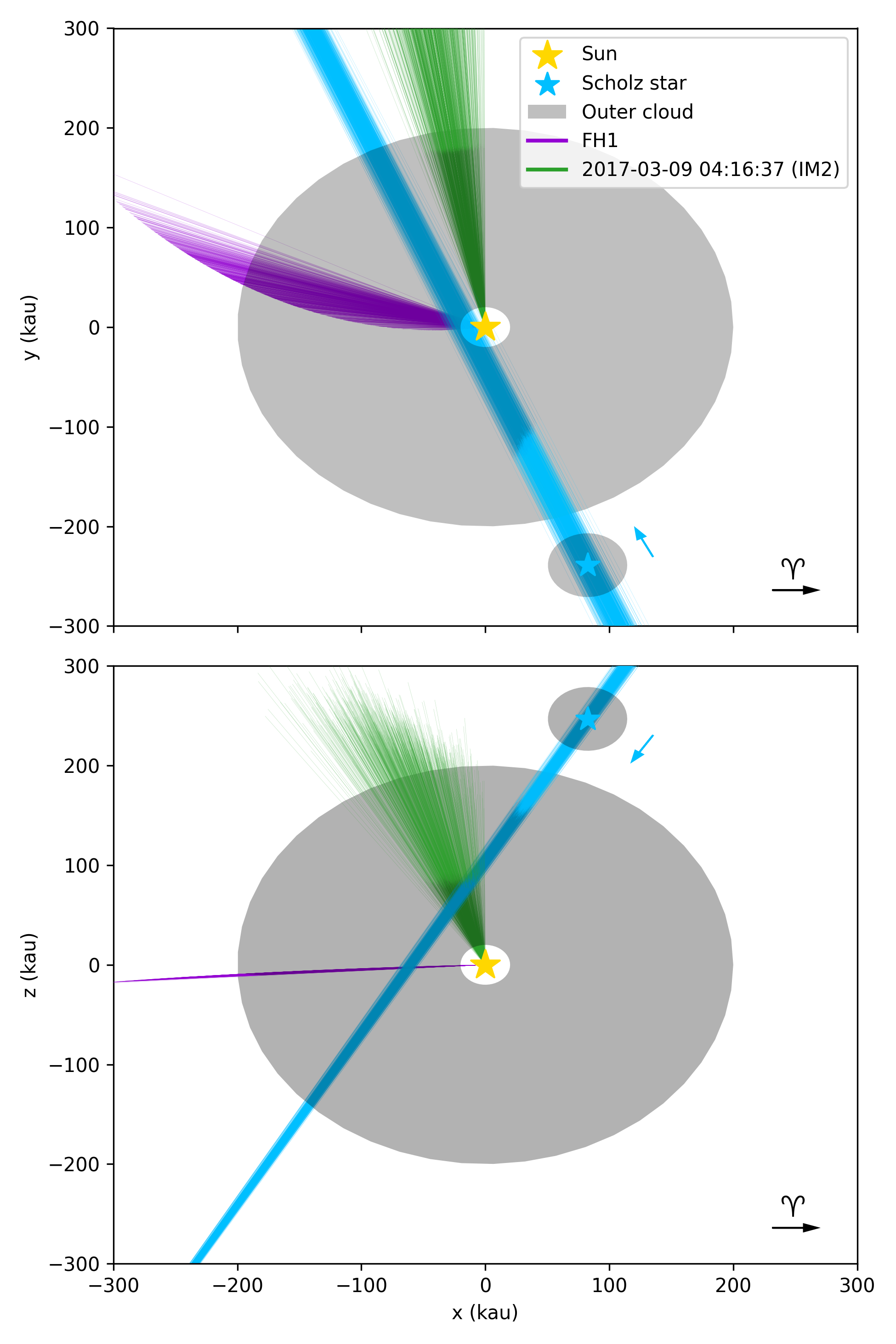}
\caption{Integration of clones for FH1, IM2, and Scholz. The diagram illustrates the heliocentric cartesian coordinates evolution during the encounter, including the respective outer clouds of material associated with both the Sun and Scholz. The arrow at the bottom right shows the direction to the point of the vernal equinox.
\label{fig:scholz_passage}}
\end{figure}

The meteoroid FH1 was 36$\pm$18 kya ago at $\sim$67 kau, and the IM2 object reached the same distance 14$\pm$2 kya ago. The closest encounter found in the simulation of FH1 with the Scholz trajectory was at 39 kau, while IM2 passed at 131 kau. In spite of almost intersecting trajectories, the excess velocity of IM2 at impact (-8 km$\,$s$^{-1}$) causes it not to be compatible in time with the Scholz passage. Looking at Table \ref{tab:CNEOSdeviations}, it can be seen that $\sim$18\% of the events have velocity errors around 30\% or more of the nominal value. If the IM2 measurement had an uncertainty of 22\% it would be perfectly compatible in time with the Scholz fly-by. Velocities proximal to the parabolic limit for FH1 are consistent, both temporally and directionally, with the passage of Scholz. This consistency necessitates only a 1\% reduction in Scholz's nominal velocity, a value well within the estimated range of uncertainty. Figure \ref{fig:scholz_passage} presents the geometric configuration of the encounter involving FH1, IM2, and Scholz.

Long-period objects can acquire excess velocity from relatively low gravitational perturbations with no need for very close encounters if they are oriented in the appropriate direction at the appropriate time. Moreover, the perturbation may not necessarily have occurred during the time of the maximum approach of Scholz, which took $\sim$21.5 kya to traverse the Oort cloud. Considering that an object is fixed in reference to the solar system barycenter, it is possible to estimate the time-integrated impulse exerted by a passing star from classical impulse approximation \citep{Rickman2005EMP}:
\begin{equation}\label{eq:DV}
\Delta v = \frac{2GM_*}{v^*}\left ( \frac{\hat{b}_o}{b_o} - \frac{\hat{b}_*}{b_*} \right ),
\end{equation}

where $G$ is the gravitational constant, $b_o$ is the vector from the Oort cloud object to the Scholz closest approach, and $b_*$ is the vector from the Sun to the Scholz closest approach. 

To elucidate, consider a notional object situated at a radial distance of 39 kau beyond the point of closest approach between the Scholz star system and the Sun, which occurs at 68.7 kau. According to Eq. \ref{eq:DV}, the interaction with the Scholz could impart a maximum velocity change of approximately 0.136 m$\,$s$^{-1}$. In the defined encounter geometry within the Oort Cloud, the object would need to achieve a velocity of 0.982 m$\,$s$^{-1}$ to attain a perihelion distance of 1 au. The velocity impulse generated by the Scholz star would represent approximately 14\% of the object's perigee velocity. Therefore, such a perturbation has the capacity to significantly modify the orbital parameters of a distantly located object, resulting in a highly eccentric inbound trajectory toward Earth.

Detection of exocomets and warm and cold debris belts around stars suggests the existence of Oort-like outer clouds material in other stellar systems \citep{Marois2010Natur, Kiefer2014Natur}. Highly eccentric evaporating comets are compatible with the metallic absorption lines observed in debris disk spectra, which may be evidence of exocomet clouds \citep{Beust1990AA, Hanse2018MNRAS}. \citet{Hanse2018MNRAS} found that 25\% to 65\% of the mass lost from the Oort cloud is due to objects being either injected into the planetary region or ejected into interstellar space mainly because of stellar encounters. A hypothetical Oort-like Scholz cloud should have the outer edge less distant from its center than the Sun due to its smaller mass. Assuming that both stellar systems have undergone similar processes, we can establish that the binding energies of their outer clouds of material are roughly the same and, therefore, their gravitational potential energies as well:

\begin{equation}
 -\frac{G M_{\odot} m_o}{R_{\odot}} = -\frac{G M_* m_o}{R_*},
\end{equation}

where $R$ is the distance from the objects to the star, $m_o$ is the mass of the surrounding objects, $\odot$ subscript refers to the Sun, and $*$ to another star. Accordingly, the outer edge of the Scholz outer cloud should be scaled down as $R_*=0.16R_\odot$.

Given that the classical outer edge of the Oort cloud is 200 kau \citep{Dones2004}, it is expected that the Scholz system has an outer cloud edge of 32 kau, close enough to the Sun for any object to be disturbed towards the planetary region during the fly-by, even more so when Scholz Hill sphere at the closest approach was 26 kau. In fact, IM2 clone trajectory traversed the path of the Scholz cloud and a considerable percentage of the clones intersected its Hill sphere path, while some FH1 clones passed at 2 Scholz Hill sphere radius. These possible injections could be facilitated by the joined effect of the Galactic disc, other passing stars interactions, and the presence of massive objects on the outskirts of the Scholz system or of our solar system. For example, the Scholz motion region passes through the zone of high probability for the putative planet 9 and multiple hyperbolic CNEOS fireballs fall around it \citep{Batygin2016AJ, BrownBatygin2021AJ, HectorSocas2022arXiv}. Note that although the maximum velocity change in a fly-by is double the relative velocity of the encounter, a very massive nearly static object could redirect another with a zero net velocity change into a hyperbolic orbit due to the new geometry of the motion with respect to the central body. However, this situation would lead to slow unbound orbits, which might explain the modest impact velocity excess for many hyperbolic solar objects.

With successive data releases from Gaia, the space observatory of the European Space Agency, the identification of new stellar close encounters has been increasing. For example, in the first Gaia data release (GDR1) 2 stars (out of 300,000) were found to come within 1 pc of the Sun \citep{Bailer2018AA}, while in GDR2 were 26 stars (out of 7.2 million) passing within the same distance \citep{Baileretal2018AA} and 61 stars (out of 33 million) in GDR3 \citep{Bailer2022ApJ}. \citet{Baileretal2018AA} inferred the present rate of stellar encounters to be 19.7$\pm$2.2 per Myr within 1 pc. This implies that the Oort cloud is expected to have experienced $\sim$2 interloper visits in the last 80 kya. However, only one has been identified during this period of time.

We stress that as uncertainties of CNEOS data are unavailable, no definite conclusions can be established regarding IM2. Nevertheless, considering that there may be complex gravitational interactions, we claim that both FH1 and IM2 are consistent with being gravitationally accelerated impactors originating from the Oort Cloud, likely injected during Scholz's recent fly-by. Additionally, IM2 could plausibly be an object ejected from the outer cloud of the Scholz binary system. The other CNEOS hyperbolic events (if they are not outliers) possibly have experienced close encounters with massive objects (such as stars, free-floating brown dwarfs, rogue planets, sub-stellar or sub-Jovian mass perturbers, rogue planets, primordial black holes...) when traversing the Oort cloud less time ago than the Scholz passage, which in the case of a star could be supported by the current rate of stellar encounters.

\section{Conclusions}

We have analyzed an unusual grazing meteor (FH1) recorded in October 2022 by the Finnish Fireball Network. The cm-sized meteoroid exhibited an inbound likely hyperbolic orbit and an (at least) asteroidal consistency. Considering that its orbital plane coincides with the ecliptic and is close to the parabolic velocity limit, it seems more likely to be a perturbed Oort cloud object rather than an interstellar interloper. Within the estimated uncertainties, FH1 could be associated with the known most recent stellar encounter with our solar system, i.e., the Scholz system.

4 of the 6 hyperbolic CNEOS fireballs exhibit a statistical oddity in the geocentric radiant distribution around the Gemini constellation, an area with an overdensity of hyperbolic radiants and identified as compatible in time with the Scholz fly-by. We show statistical evidence that these events cannot really pertain to a randomly incoming interstellar population as the likelihood of their low orbital inclinations is extremely improbable compared to the expected one (with a probability of having occurred by chance of 0.00007\%). Therefore, these impactors most likely belonged to our solar nebula and have been perturbed by massive objects lying on or intersecting the plane of the ecliptic. These massive objects could also form part of the Oort cloud, although this would limit the excess velocity of the projectiles.

Given the new mean uncertainties estimated in this work for CNEOS detections by benchmarking with 17 independent ground-based observations, the 2017-03-09 04:16:37 (IM2) fireball appears to be consistent with the Scholz close encounter if the velocity was overestimated by 22\%, which is within the error range for $\sim$18\% of events compared. In addition, IM2 showed a peak power in its light curve that corresponded to a dynamic pressure (i.e. aerodynamic strength) typical of iron meteorites, about $\sim$75 MPa, and most likely produced a recoverable metallic meteorite \citep{Eloy2022AJ, Siraj2022ApJb}. This would inaugurate a window of opportunity for stellar archaeology sample collection with known trajectory beyond the tiny presolar grains embedded in meteorites.


If these hyperbolic impactors were interstellar visitors, it would have significant implications both for the incoming flux of extra-solar objects and for the characterization of their physical properties, which would be biased toward high-strength compositions and low inclinations. If they were Oort cloud objects, FH1 would be the second cm-sized asteroid-like observed object after \citet{Vida2022NatAs} and IM2 the first detected iron-like body from the outermost part of our solar or another stellar system. This would provide further evidence for the massive proto-asteroid belt and Jupiter's "Grand Tack" dynamical instability scenario \citep{Shannon2019MNRAS}. It would imply that the Oort cloud could currently be populated not only by weak cometary objects but also by ice-free rocky material scattered by giant planets's round trip to inner orbits. 

The absence of interstellar meteorites and the low orbital inclinations of the hyperbolic projectiles studied indicate that the population of massive objects forming and crossing the Oort cloud may be larger than previously thought, injecting large meteoroids into the planetary regions. Our results reinforce the idea that passing stars or other massive objects represent a source of hyperbolic Earth impactors that must be examined in detail on a case-by-case basis before claiming the interstellar origin of any object with excess velocity.

\section*{Author Contributions}

    EP-A performed the analysis of FH1, the statistical study of CNEOS hyperbolic events, the investigation of the massive object close encounters hypothesis, and wrote the manuscript. JV coordinated the Finnish Fireball Network efforts and supported the astrometry and atmospheric trajectory calculation of FH1. JMT-R oversaw the research activity and provided scientific insights. HS-N performed the N-body simulations and suggested the possible association of FH1 with Scholz's star. MG clarified the issues with the dynamic model fit and contributed to the estimation of the terminal velocity. MS first identified FH1 as an interstellar candidate. JMT-R and AR acquired financial support for the project leading to this publication and supervised the work. All authors read, edited, and approved the manuscript.

\section*{Acknowledgements}
      This project has received funding from the European Research Council (ERC) under the European Union’s Horizon 2020 research and innovation programme (grant agreement No. 865657) for the project “Quantum Chemistry on Interstellar Grains” (QUANTUMGRAIN). JMT-R and EP-A. acknowledge financial support from the project PID2021-128062NB-I00 funded by MCIN/AEI/10.13039/501100011033. AR acknowledges financial support from the FEDER/Ministerio de Ciencia e Innovación – Agencia Estatal de Investigación (PID2021-126427NB-I00, PI: AR). MG acknowledges the Academy of Finland project no. 325806 (PlanetS). HS-N acknowledges support from the Agencia Estatal de Investigación del Ministerio de Ciencia e Innovación (AEI-MCINN) under grant Hydrated Minerals and Organic Compounds in Primitive Asteroids with reference PID2020-120464GB-I100. This work was also partially supported by the program Unidad de Excelencia María de Maeztu CEX2020-001058-M. We thank all FFN station operators and photographers whose continuous dedication has allowed to record this grazing meteor: Jarmo Moilanen, Harri Kiiskinen, Markku Lintinen, Jari Luomanen, and Kari Haila. We also thank Marc Corretgé-Gilart for his support in astrometric calibrations.

\bibliographystyle{elsarticle-harv} 
\bibliography{cas-refs}






\section*{Appendix}

    \begin{table*}[htb!]
    \centering
    \footnotesize
    \caption{Position vectors of the initial and final points of FH1's luminous path in the Earth-centered Earth-fixed coordinate system, as recorded by each of the four stations.}
    \label{tabA:ECEF_beg_end}
    \begin{tabular}{lcccccc}
    \hline
              & Nyrola & Vaala & Tampere & Sastamala  \\
    \hline
    X$_{beg}$ (m) & 2,684,184 $\pm$ 6 & 2,637,480 $\pm$ 4 & 2,743,469 $\pm$ 4 & 2,701,116 $\pm$ 3 \\
    Y$_{beg}$ (m) & 1,161,276 $\pm$ 14 & 1,188,673 $\pm$ 56 & 1,126,497 $\pm$ 7 & 1,151,343 $\pm$ 11 \\
    Z$_{beg}$ (m) & 5,787,091 $\pm$ 29 & 5,806,802 $\pm$ 26 & 5,762,070 $\pm$ 20 & 5,779,945 $\pm$ 31 \\
    X$_{end}$ (m) & 2,969,352 $\pm$ 17 & 2,891,289 $\pm$ 15 & 2,838,781 $\pm$ 2 & 2,800,918 $\pm$ 3 \\
    Y$_{end}$ (m) & 993,989 $\pm$ 114 & 1,039,782 $\pm$ 122 & 1,070,585 $\pm$ 21 & 1,092,797 $\pm$ 10 \\
    Z$_{end}$ (m) & 5,666,739 $\pm$ 24 & 5,699,684 $\pm$ 6 & 5,721,845 $\pm$ 21 & 5,737,825 $\pm$ 19 \\
    \hline
    \end{tabular}
    \end{table*}

\begin{table*}[htb!]
    \centering
    \footnotesize
    \caption{Detected positions of FH1 in a horizontal coordinate system for each frame from Nyrola. The table presents the azimuth and elevation coordinates along with their corresponding frame numbers. The mean residuals from astrometry are 0.0139 $^\circ$ for azimuth and 0.0121 $^\circ$ for elevation.}
    \label{tabA:NyrolaLocalObs}
    \begin{tabular}{ccc|ccc|ccc}
        \hline
        \# & Azi. (º) & Elev. (º) & \# & Azi. (º) & Elev. (º) & \# & Azi. (º) & Elev. (º) \\
        \hline
11 & 317.3931 & 37.8781 & 51 & 281.3995 & 29.5673 & 91 & 263.1464 & 20.2415 \\
12 & 316.2749 & 37.8103 & 52 & 280.7005 & 29.2648 & 92 & 262.8051 & 20.0648 \\
13 & 315.3540 & 37.7320 & 53 & 280.0772 & 28.9837 & 93 & 262.5361 & 19.8995 \\
14 & 314.1403 & 37.6221 & 54 & 279.4517 & 28.7645 & 94 & 262.2554 & 19.7190 \\
15 & 313.1072 & 37.5227 & 55 & 278.8840 & 28.4955 & 95 & 261.9023 & 19.5021 \\
16 & 311.9602 & 37.4195 & 56 & 278.2156 & 28.2280 & 96 & 261.6241 & 19.3508 \\
17 & 310.8714 & 37.2770 & 57 & 277.6835 & 27.9869 & 97 & 261.3567 & 19.1678 \\
18 & 309.8847 & 37.1485 & 58 & 277.0828 & 27.6925 & 98 & 261.1147 & 19.0138 \\
19 & 308.7614 & 36.9825 & 59 & 276.4748 & 27.4461 & 99 & 260.8124 & 18.8077 \\
20 & 307.8117 & 36.8475 & 60 & 276.0162 & 27.2016 & 100 & 260.5358 & 18.6459 \\
21 & 306.6819 & 36.6881 & 61 & 275.4985 & 26.9515 & 101 & 260.2954 & 18.4980 \\
22 & 305.7508 & 36.5257 & 62 & 274.9191 & 26.6835 & 102 & 260.0438 & 18.3728 \\
23 & 304.6602 & 36.3582 & 63 & 274.3874 & 26.4317 & 103 & 259.7673 & 18.1635 \\
24 & 303.5926 & 36.1520 & 64 & 273.8917 & 26.1857 & 104 & 259.4929 & 18.0146 \\
25 & 302.6033 & 35.9439 & 65 & 273.3587 & 25.9597 & 105 & 259.2896 & 17.8871 \\
26 & 301.5501 & 35.7376 & 66 & 272.8778 & 25.6810 & 106 & 259.0386 & 17.7289 \\
27 & 300.6744 & 35.5661 & 67 & 272.3491 & 25.4399 & 107 & 258.8114 & 17.5625 \\
28 & 299.6912 & 35.3705 & 68 & 271.8988 & 25.2108 & 108 & 258.5265 & 17.3885 \\
29 & 298.7177 & 35.1155 & 69 & 271.4487 & 24.9688 & 109 & 258.3008 & 17.2594 \\
30 & 297.7942 & 34.9211 & 70 & 271.0233 & 24.7566 & 110 & 258.0748 & 17.0994 \\
31 & 296.9946 & 34.6887 & 71 & 270.5782 & 24.5114 & 111 & 257.8603 & 16.9311 \\
32 & 295.9840 & 34.4713 & 72 & 270.0966 & 24.2465 & 112 & 257.6202 & 16.7871 \\
33 & 295.1154 & 34.1997 & 73 & 269.6739 & 23.9963 & 113 & 257.3981 & 16.6365 \\
34 & 294.3219 & 34.0078 & 74 & 269.2575 & 23.7892 & 114 & 257.1762 & 16.4860 \\
35 & 293.4068 & 33.7414 & 75 & 268.8598 & 23.5792 & 115 & 256.9630 & 16.3234 \\
36 & 292.5561 & 33.4805 & 76 & 268.4249 & 23.3262 & 116 & 256.7421 & 16.2062 \\
37 & 291.7253 & 33.2369 & 77 & 268.0482 & 23.1121 & 117 & 256.5569 & 16.0777 \\
38 & 290.9343 & 33.0065 & 78 & 267.6577 & 22.9201 & 118 & 256.3266 & 15.9198 \\
39 & 290.0894 & 32.7342 & 79 & 267.2421 & 22.6909 & 119 & 256.1412 & 15.8074 \\
40 & 289.3101 & 32.4818 & 80 & 266.8870 & 22.4885 & 120 & 255.9655 & 15.7245 \\
41 & 288.5378 & 32.2250 & 81 & 266.4824 & 22.2579 & 121 & 255.7539 & 15.5943 \\
42 & 287.7816 & 31.9863 & 82 & 266.1553 & 22.0821 & 122 & 255.5601 & 15.4522 \\
43 & 287.0466 & 31.7155 & 83 & 265.7647 & 21.8330 & 123 & 255.3926 & 15.2929 \\
44 & 286.1926 & 31.4039 & 84 & 265.4642 & 21.6612 & 124 & 255.2166 & 15.1914 \\
45 & 285.4432 & 31.1795 & 85 & 265.1273 & 21.4680 & 125 & 255.0321 & 15.0947 \\
46 & 284.7507 & 30.8916 & 86 & 264.7417 & 21.2385 & 126 & 254.8746 & 14.9874 \\
47 & 284.0495 & 30.6277 & 87 & 264.4171 & 21.0196 & 127 & 254.7251 & 14.8853 \\
48 & 283.3727 & 30.3440 & 88 & 264.1204 & 20.8522 & 128 & 254.5671 & 14.7714 \\
49 & 282.7171 & 30.1159 & 89 & 263.7876 & 20.6626 & 129 & 254.4091 & 14.6573 \\
50 & 281.9793 & 29.8009 & 90 & 263.4555 & 20.4567 & 130 & 254.2257 & 14.5249 \\
        \hline
    \end{tabular}
\end{table*}

\begin{table*}[htb!]
    \centering
    \footnotesize
    \caption{Detected positions of FH1 in a horizontal coordinate system for each frame from Vaala. The table presents the azimuth and elevation coordinates along with their corresponding frame numbers. The mean residuals from astrometry are 0.0197 $^\circ$ for azimuth and 0.0227 $^\circ$ for elevation.}
    \label{tabA:VaalaLocalObs}
    \begin{tabular}{ccc|ccc|ccc}
        \hline
        \# & Azi. (º) & Elev. (º) & \# & Azi. (º) & Elev. (º) & \# & Azi. (º) & Elev. (º) \\
        \hline
1 & 236.5024 & 40.6797 & 37 & 234.6534 & 24.9055 & 73 & 233.8727 & 16.7750 \\
2 & 236.4637 & 40.1331 & 38 & 234.6508 & 24.5810 & 74 & 233.8427 & 16.5849 \\
3 & 236.4142 & 39.6152 & 39 & 234.6147 & 24.2971 & 75 & 233.8378 & 16.4120 \\
4 & 236.3346 & 38.9983 & 40 & 234.5789 & 24.0129 & 76 & 233.8159 & 16.2693 \\
5 & 236.2319 & 38.4367 & 41 & 234.5689 & 23.7488 & 77 & 233.8006 & 16.0920 \\
6 & 236.2106 & 37.8006 & 42 & 234.5588 & 23.4434 & 78 & 233.8024 & 15.9098 \\
7 & 236.0623 & 37.3039 & 43 & 234.5294 & 23.1309 & 79 & 233.8085 & 15.7675 \\
8 & 236.0186 & 36.7238 & 44 & 234.4965 & 22.8900 & 80 & 233.7928 & 15.6158 \\
9 & 235.9471 & 36.2442 & 45 & 234.4648 & 22.6491 & 81 & 233.7849 & 15.4864 \\
10 & 235.9124 & 35.7728 & 46 & 234.4474 & 22.4038 & 82 & 233.7755 & 15.3248 \\
11 & 235.8074 & 35.1764 & 47 & 234.4247 & 22.1466 & 83 & 233.7612 & 15.1485 \\
12 & 235.7534 & 34.7154 & 48 & 234.3921 & 21.8638 & 84 & 233.7480 & 15.0289 \\
13 & 235.7152 & 34.2276 & 49 & 234.3605 & 21.6066 & 85 & 233.7318 & 14.9025 \\
14 & 235.6506 & 33.7938 & 50 & 234.3589 & 21.3495 & 86 & 233.7164 & 14.7759 \\
15 & 235.5591 & 33.3065 & 51 & 234.3200 & 21.1474 & 87 & 233.6986 & 14.6423 \\
16 & 235.5453 & 32.8562 & 52 & 234.2690 & 20.9154 & 88 & 233.7021 & 14.5191 \\
17 & 235.5569 & 32.3975 & 53 & 234.2688 & 20.7354 & 89 & 233.6894 & 14.3999 \\
18 & 235.4853 & 31.9917 & 54 & 234.2627 & 20.5153 & 90 & 233.6885 & 14.2619 \\
19 & 235.4064 & 31.5686 & 55 & 234.2130 & 20.3090 & 91 & 233.6790 & 14.1260 \\
20 & 235.3152 & 31.0649 & 56 & 234.2046 & 20.0306 & 92 & 233.6814 & 13.9712 \\
21 & 235.2724 & 30.6519 & 57 & 234.1940 & 19.7954 & 93 & 233.6727 & 13.8354 \\
22 & 235.2357 & 30.1940 & 58 & 234.2006 & 19.6316 & 94 & 233.6600 & 13.7175 \\
23 & 235.1817 & 29.8083 & 59 & 234.1991 & 19.3949 & 95 & 233.6379 & 13.5695 \\
24 & 235.1289 & 29.4228 & 60 & 234.1542 & 19.1211 & 96 & 233.6235 & 13.4441 \\
25 & 235.0852 & 29.0555 & 61 & 234.1356 & 18.9411 & 97 & 233.6147 & 13.3094 \\
26 & 235.0264 & 28.6525 & 62 & 234.1343 & 18.7553 & 98 & 233.6142 & 13.1893 \\
27 & 235.0307 & 28.2694 & 63 & 234.0993 & 18.5305 & 99 & 233.6123 & 13.0649 \\
28 & 234.9547 & 27.9376 & 64 & 234.0808 & 18.3512 & 100 & 233.6168 & 12.9437 \\
29 & 234.9304 & 27.5458 & 65 & 234.0312 & 18.1898 & 101 & 233.6102 & 12.8553 \\
30 & 234.9257 & 27.2064 & 66 & 234.0048 & 18.0134 & 102 & 233.6139 & 12.7641 \\
31 & 234.8863 & 26.8479 & 67 & 233.9729 & 17.8215 & 103 & 233.5709 & 12.6515 \\
32 & 234.8490 & 26.5312 & 68 & 233.9511 & 17.6021 & 104 & 233.5740 & 12.5760 \\
33 & 234.7756 & 26.1715 & 69 & 233.9112 & 17.4138 & 105 & 233.5425 & 12.4799 \\
34 & 234.7705 & 25.8454 & 70 & 233.8998 & 17.2750 & 107 & 233.5334 & 12.2986 \\
35 & 234.7604 & 25.5074 & 71 & 233.8739 & 17.1247 & 108 & 233.4896 & 12.1700 \\
36 & 234.7124 & 25.2397 & 72 & 233.8495 & 16.9489 &  &  &  \\
        \hline
    \end{tabular}
\end{table*}

\end{document}